\pdfoutput=1
\pdfoutput=1
\pdfoutput=1
\pdfoutput=1
\pdfoutput=1
\documentclass[journal,10pt]{IEEEtran}
\usepackage{amssymb}
\usepackage{amsmath}
\usepackage{cite}
\usepackage{url}
\usepackage{xcolor}
\usepackage{cite,graphicx,amsmath,amssymb}
\usepackage{fancyhdr}
\usepackage{mdwmath}
\usepackage{mdwtab}
\usepackage{amsthm}
\usepackage{setspace}
\usepackage{algorithm}
\usepackage{algorithmic}
\usepackage{makecell}
\usepackage{diagbox}
\pdfoutput=1

\newtheorem{remark}{Remark}
\newtheorem{theorem}{Theorem}

\newtheorem{lemma}{Lemma}

\newtheorem{corollary}{Corollary}

\allowdisplaybreaks
\makeatletter
\def\ScaleIfNeeded{%
\ifdim\Gin@nat@width>\linewidth \linewidth \else \Gin@nat@width
\fi } \makeatother
\begin{document}
%\title{Fundamental Coverage Tradeoff of STAR-RIS Aided Communication with NOMA and OMA}
\title{Coverage Characterization of STAR-RIS Networks: NOMA and OMA}
% \markboth{\textit{A Manuscript Submitted to The IEEE Communications Magazine} }

\author{
 Chenyu Wu, Yuanwei~Liu, \IEEEmembership{Senior Member,~IEEE}, Xidong Mu, \IEEEmembership{Graduate Student Member, IEEE}, Xuemai Gu, \IEEEmembership{Member,~IEEE}, Octavia A. Dobre, \IEEEmembership{Fellow,~IEEE}\vspace{-2em}

%\IEEEauthorblockN{,~\IEEEmembership{Member,~IEEE,}}
% ,~\IEEEmembership{Fellow,~IEEE,}
%and \IEEEauthorblockN{,~\IEEEmembership{Senior Member,~IEEE,}

%\thanks{Y. Liu and X. Liu are with Queen Mary University of London; X. Mu is with Beijing University of Posts and Telecommunications; M. Di. Renzo is with Universit\'e Paris-Saclay; Z. Ding is with the The University of Manchester; R. Schober is with Friedrich-Alexander-University Erlangen-N{\"u}rnberg (FAU).}
%\thanks{\emph{Corresponding author: C. Wu.}}
\thanks{C. Wu, and X. Gu are with the School of Electronic and Information Engineering, Harbin Institute of Technology (HIT), Harbin, 150001, China. (e-mail: \{wuchenyu, guxuemai\}@hit.edu.cn).}
\thanks{Y. Liu is with the School of Electronic Engineering and Computer Science, Queen Mary University of London, London E1 4NS, UK, (email: yuanwei.liu@qmul.ac.uk).}
\thanks{X. Mu is with School of Artificial Intelligence, Beijing University of Posts and Telecommunications, Beijing, 100876, China (email: muxidong@bupt.edu.cn).}
\thanks{O. A. Dobre is with the Department of Electrical and Computer Engineering, Memorial University, St. John’s, NL A1C 5S7, Canada (e-mail: odobre@mun.ca).}
}

\maketitle

%\begin{abstract}
%This article focuses on the exploitation of reconfigurable intelligent surfaces (RISs) in multi-user networks employing orthogonal multiple access (OMA) or non-orthogonal multiple access (NOMA), with an emphasis on investigating the interplay between NOMA and RIS. Depending on whether the RIS reflection coefficients can be adjusted only once or multiple times during one transmission, we distinguish between \emph{static} and \emph{dynamic} RIS configurations. In particular, the capacity region of RIS aided single-antenna NOMA networks is characterized and compared with the OMA rate region from an information-theoretic perspective, revealing that the dynamic RIS configuration is capacity-achieving. Then, the impact of the RIS deployment location on the performance of different multiple access schemes is investigated, which reveals that asymmetric and symmetric deployment strategies are preferable for NOMA and OMA, respectively. Furthermore, for RIS aided multiple-antenna NOMA networks, three novel joint active and passive beamformer designs are proposed based on both beamformer based and cluster based strategies. Finally, open research problems for RIS-NOMA networks are highlighted.
%\end{abstract}
%\begin{IEEEkeywords}
%Heterogeneous ultra dense networks, non-orthogonal multiple access, massive connectivity, user association, and resource allocation.
%\end{IEEEkeywords}
\begin{abstract}
The novel concept of simultaneously transmitting and reflecting reconfigurable intelligent surface (STAR-RIS) is investigated, where incident signals can be transmitted and reflected to users located at different sides of the surface. In particular, the fundamental coverage range of STAR-RIS aided two-user communication networks is studied. A sum coverage range maximization problem is formulated for both non-orthogonal multiple access (NOMA) and orthogonal multiple access (OMA), where the resource allocation at the access point and the transmission and reflection coefficients at the STAR-RIS are jointly optimized to satisfy the communication requirements of users. For NOMA, we transform the non-convex decoding order constraint into a linear constraint and the resulting problem is convex, which can be optimally solved. %by jointly optimizing the power allocation scheme and transmission and reflection coefficients. 
%The problem is solved by jointly optimizing the network resource allocation and transmission and reflection coefficients. 
%We further study the general energy splitting (ES) case, where the coefficients of each element can be modulated separately. 
For OMA, we first show that the optimization problem for given time/frequency resource allocation is convex. Then, we employ the one dimensional search-based algorithm to obtain the optimal solution.
Numerical results %demonstrate that NOMA outperforms OMA in terms of coverage range for STAR-RIS. 
%Compared with two baselines, namely fixed coefficient and conventional RIS, the STAR-RIS could extend the coverage range significantly.
%In addition, the superiority of ES is more obvious when the channel fading is dominant.
%It also 
reveal that the coverage can be significantly extended by the STAR-RIS compared with conventional RISs.
\end{abstract}
~\\
\section{Introduction}
Reconfigurable intelligent surfaces (RISs) have been envisioned as a revolutionary technology to enhance the spectrum efficiency (SE) and to improve the coverage range for beyond fifth-generation (B5G) wireless communication networks\cite{jsacris,RIS2}. An RIS is composed of massive low-cost and programmable elements, and thus, can reconfigure the propagation of incident wireless signals by adjusting the amplitudes and phase shifts of each element. %Significant performance enhancement is achieved with lower energy consumption by RIS due to the passive reflection compared with active relaying.
Due to the nearly passive working mode, RISs can enhance the communication performance without the need of radio frequency (RF) chains compared with active relaying, which reduces the energy consumption and the hardware costs \cite{access}. Moreover, when the direct link between the access point (AP) and users is blocked, RIS can be deployed to provide additional signal paths, thus satisfying the basic communication requirements of users in the signal dead zone. 
Driven by the above advantages, RISs have received extensive interests from both industry and academy to fully exploit its benefits. 
%Recently, some research works explore the interplay between RIS and non-orthogonal multiple access (NOMA), which further boosts the performance of wireless communication networks\cite{mu1,zr1,ni1}. 
The adoption and superiority of RISs for communication networks have been studied in previous works\cite{irs2,mu1,irs1,oct1,oct2}, where the joint beamforming optimization and the energy efficiency analysis have been investigated.
However, these contributions mainly focus on RISs which act as reflective metasurfaces; hence, the served users ought to be on the same side of RIS, i.e. \emph{half-space coverage}, which limits the flexibility of deploying RISs.

%Non-orthogonal multiple access (NOMA), as a emerging multiple access technique, will play incredibly important roles in providing massive connection, guaranteeing access fairness, and improving SE in future wireless networks\cite{surveynoma,proceeding}. In power domain NOMA, the desired signals for different users can be transmitted over the same time-frequency blocks by utilizing superposition coding and successive interference cancellation at the transmitter and receiver, respectively\cite{powerdomain}. Correspondingly, new challenges are also brought by this novel framework, such as the designs of network resource allocation strategy and decoding orders of different users. Some prior works have explored particularity of NOMA compared with orthogonal multiple access (OMA) in RIS-assisted scenario\cite{mu1,zr1,ni1}. 

To overcome this drawback, recently, a novel concept of simultaneous transmitting and reflecting RISs (STAR-RISs) has been proposed\cite{SRAR,STAR}. Different from existing reflecting-only RISs, STAR-RISs can simultaneously transmit and reflect the incident signals. Hence, \emph{full-space coverage} can be enabled. STAR-RISs provide new degree-of-freedom for manipulating signal propagation, thus increasing the flexibility for network design. Despite the above appealing characteristics, to the best of our knowledge, the superiority of STAR-RISs in terms of coverage range has not been studied yet, which motivates this work.
%In particular, to integrate STAR-RIS to future wireless network, the fundamental performance comparison between different multiple access schemes incorporating STAR-RIS is essential.
%fundamental coverage analysis of STAR-RIS-assisted multiple access (MA) is not well studied. 

In this article, we aim to characterize the fundamental coverage range of STAR-RIS aided communication networks. In particular, an AP communicates with one transmitted user (T user) and one reflected user (R user) employing both non-orthogonal multiple access (NOMA) and orthogonal multiple access (OMA) with the aid of an STAR-RIS. %We present the STAR-RIS model via transmission and reflection, as well as multiple access schemes including NOMA and OMA. Then, 
A sum coverage range maximization problem for the joint optimization of the resource allocation at the AP and the transmission and reflecting coefficients at the STAR-RIS is formulated for each multiple access scheme, subject to the quality-of-service (QoS) requirements of users.
%under NOMA and OMA are formulated and solved by jointly optimizing the network resource allocation and transmission and reflection coefficients. 
Specifically, for NOMA, we convert the joint optimization problem into a convex one by transforming the non-convex decoding order constraint. For OMA, we present a convex subproblem with given time/frequency resource allocation and then employ the one dimensional search-based algorithm to obtain the optimal solution.
%prove that problem for given frequency/time resource allocation is convex, which can be efficiently solved by existing solvers. Then, we can use one-dimensional bisection search for optimal solutions with low complexity. 
Numerical results unveil that the coverage range of STAR-RISs is significantly enhanced by NOMA. Moreover, compared with conventional RIS, STAR-RIS can greatly extend the network coverage range for both NOMA and OMA.

The remainder of this article is organized as follows. The system model and problem formulation are presented in Section II. Efficient algorithms are designed to solve problems for NOMA and OMA in Section III. In Section IV, numerical results are presented to verify the effectiveness of STAR-RISs. Finally, Section V concludes this article.

%and analyze its performance in terms of coverage range. we depict the Pareto boundary of coverage range for STAR-RIS with different MA scheme, i.e, equal time/frequency division OMA (OMA-I), adaptive OMA (OMA-II) and NOMA. A coverage range maximization problem is formulated using distance profile and solved.

% Specifically, We first prove that the problems of OMA-I and NOMA regardless of the decoding order are convex, which can be efficiently solved by existing solvers. We then propose a successive convex approximation (SCA)-based algorithm to tackle the non-convexity of OMA-II and a linear relaxation method for NOMA. We further extend to a general energy splitting (ES) case where each element's transmission and reflection coefficients can be modulated separately. Numerical results unveil that the coverage distance of NOMA is always larger than that of OMA and the performance gain of ES is more prominent when there are less Line-of-Sight (LoS) channels. In conclusion, the proposed STAR-RIS framework supports wider coverage and provides new degrees-of-freedom (DoF) and flexibility for network design.

\section{System Model and Problem Formulation}
\subsection{System Model}
We consider a narrow-band STAR-RIS aided downlink communication network operating over frequency-flat channels, where a single-antenna AP communicates with two single-antenna users with the aid of an STAR-RIS consisting of $M$ elements. As illustrated in Fig. 1, the T user is located behind the STAR-RIS (i.e., transmission region), while the R user is located in front of the STAR-RIS (i.e., reflection region). We assume that the direct communication links between the AP and the two users are blocked by obstacles. Let $\mathbf{r}_k\in\mathbb{C}^{M\times1}$ denote the channel between STAR-RIS and user $k\in\mathcal{K}=\{t,r\}$, which is modeled as the following Rician fading channel:
\begin{figure}[t!]
	\centering
	\includegraphics[width=3in]{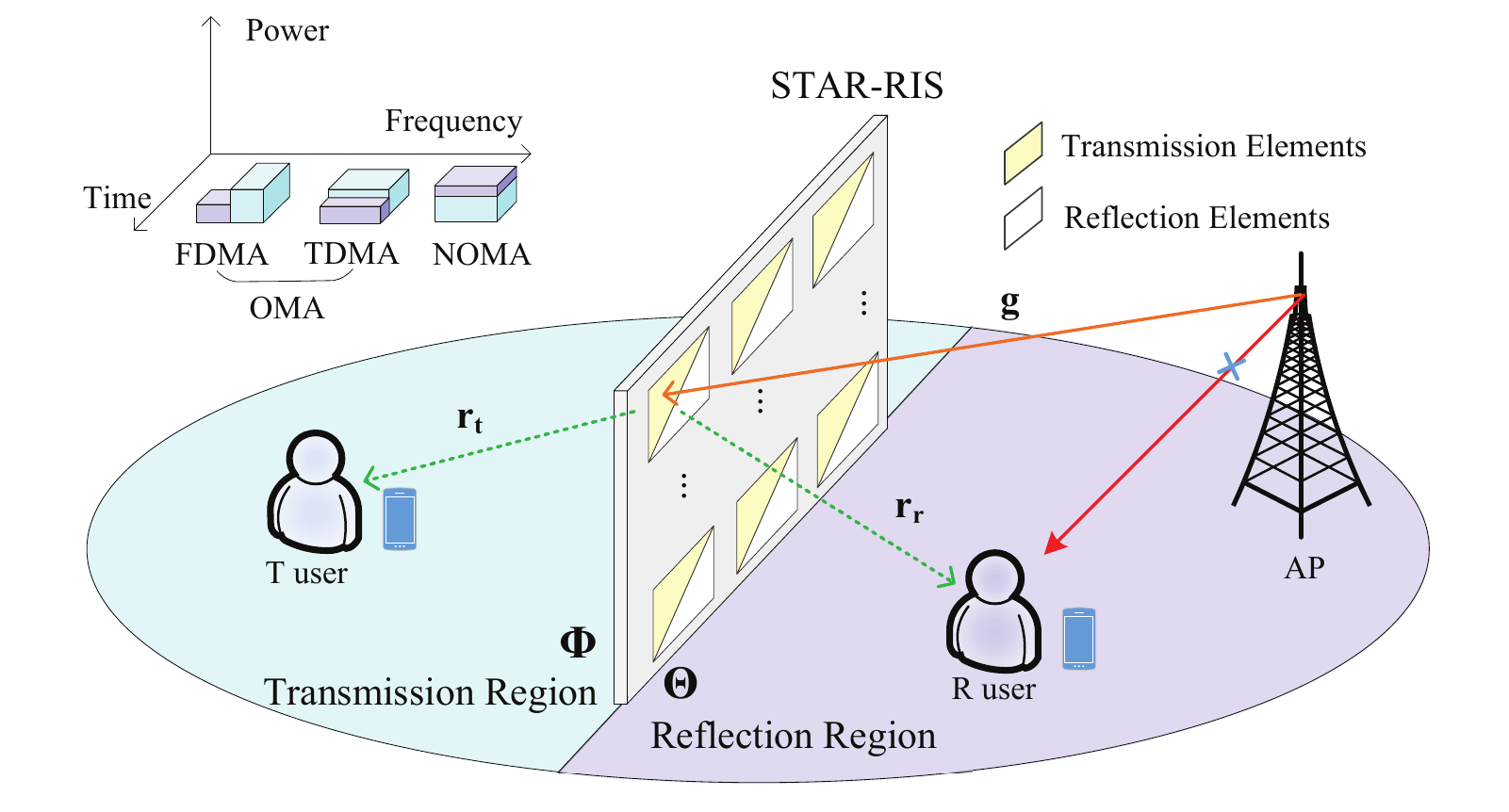}\\
	\caption{Illustration of an STAR-RIS aided two-user downlink communication systems employing NOMA and OMA.}
\end{figure}
\begin{align}\label{reflection channel}
{{\mathbf{r}}_k} = \sqrt {\frac{{{\rho _0}}}{{D_k^{{\alpha _{RU}}}}}} \Big( {\underbrace {\sqrt {\frac{{{K_{RU}}}}{{{K_{RU}} + 1}}} {\mathbf{r}}_k^{{\rm{LoS}}} + \sqrt {\frac{1}{{{K_{RU}} + 1}}} {\mathbf{r}}_k^{{\rm{NLoS}}}}_{{{\overline {\mathbf{r}} }_k}}} \Big),
\end{align}
where ${D_k}$ denotes the distance between the STAR-RIS and user $k$, ${{\alpha _{RU}}}\geq2$ denotes the path loss exponent, ${\rho _0}$ represents the path loss at a reference distance of 1 meter, ${{\mathbf{r}}_k^{{\rm{LoS}}}}\in\mathbb{C}^{M\times1}$ and ${{\mathbf{r}}_k^{{\rm{NLoS}}}}\in\mathbb{C}^{M\times1}$ are the deterministic line-of-sight (LoS) component of the array response and the random non-line-of-sight (NLoS) component modeled as Rayleigh fading, respectively, $K_{{RU}}$ denotes the Rician factor. For simplicity, we use $\overline{\mathbf{r}}_k$ to represent the summed terms of ${{\mathbf{r}}_k^{{\rm{LoS}}}}$ and ${{\mathbf{r}}_k^{{\rm{NLoS}}}}$.

Similarly, the Rician fading channel of the AP-STAR-RIS link, denoted by $\mathbf{g}\in\mathbb{C}^{M\times1}$ is expressed as:
\begin{align}\label{AP RIS channel}
{\mathbf{g}} = \sqrt {\frac{{{\rho _0}}}{{{d^{{\alpha _{AR}}}}}}} \left( {\sqrt {\frac{{{K_{AR}}}}{{{K_{AR}} + 1}}} {\mathbf{g}}^{{\rm{LoS}}} + \sqrt {\frac{1}{{{K_{AR}} + 1}}} {\mathbf{g}}^{{\rm{NLoS}}}} \right),
\end{align}
where ${\mathbf{g}}^{{\rm{LoS}}}\in\mathbb{C}^{M\times1}$ and ${\mathbf{g}}^{{\rm{NLoS}}}\in\mathbb{C}^{M\times1}$ are the LoS and NLoS components, respectively. $d$ is the distance between AP and STAR-RIS, $\alpha_{AR}$ and $K_{AR}$ are path loss exponent and Rician factor, respectively.
\subsection{STAR-RIS Model}
Different from existing works\cite{irs1,irs2,mu1,oct1,oct2}, each STAR-RIS element can simultaneously operate in two modes, namely transmission and reflection\cite{STAR,SRAR}. For transmission, the STAR-RIS allows the incident signal to pass through it via reconfiguring the signal propagation to the T user. For reflection, the RIS reflects and reconfigures the incident signal propagation to the R user. In order to reduce the overhead for information exchange between the AP and the STAR-RIS, we assume that all elements have the same amplitude coefficients. Specifically, let ${\mathbf{\Theta}_t} = \sqrt {{\beta_t}} {\rm{diag}}\left( {{e^{j{\theta_1^t}}},{e^{j{\theta_2^t}}}, \ldots ,{e^{j{\theta_M^t}}}} \right)$ and ${\mathbf{\Theta}_r} =\sqrt {{\beta_r}} {\rm{diag}}\left( {{e^{j{\theta _1^r}}},{e^{j{\theta_2^r}}}, \ldots ,{e^{j{\theta _M^r}}}} \right)$ denote the STAR-RIS transmission-coefficient and reflection-coefficient matrices, respectively. In particular, $\sqrt {{\beta_t}},\sqrt {{\beta_r}}\in[0,1]$ and $\theta_m^t,\theta_m^r\in[0,2\pi),m\in\mathcal{M}=\{1,...,M\}$ characterize the amplitude and phase shift adjustments imposed on the incident signals facilitated by the $m$th element during transmission and reflection, respectively. Note that due to the law of energy conservation, the sum energy of the transmitted and reflected signals has to be equal to that of the incident signals\cite{STAR,SRAR}; then, we have $\beta_r+\beta_t=1$. 

Further, the effective channel power gain of user $k$ is given by
\begin{align}\label{Hk}
{\left| {{h_k}} \right|^2} = {\left| {{\mathbf{r}}_k^H{{\mathbf{\Theta }}_k}{\mathbf{g}}} \right|^2} = \frac{{{\rho _0}{\beta _k}}}{{D_k^{{\alpha _{RU}}}}}\left| {{{\mathbf{q}}_k}{{\mathbf{v}}_k}} \right|^2,
\end{align}
where ${{\mathbf{q}}_k} = \overline {\mathbf{r}} _k^H{\rm{diag}}\left( {\mathbf{g}} \right)=\{q_{m,k},m=1,...,M,\forall{k}\}$ and ${{\mathbf{v}}_k} \triangleq {\left[ {e^{j{\theta _1^k}},e^{j\theta _2^k}, \ldots ,e^{j\theta _M^k}} \right]^T}$ is defined as phase shift vectors for transmission and reflection. Let $p_k$ denote the transmit power allocated to user $k$. In the following, we consider NOMA and OMA transmission schemes.

\subsection{Multiple Access Schemes}
\subsubsection{NOMA} For NOMA, the AP transmits the superposition coded signals of the two users throughout the same time and frequency resources. Let $s_k$ denote the transmitted signal for user $k$ from the AP and ${\mathbb{E}}\left[ {{{\left| {{s_k}} \right|}^2}} \right] = 1$. The received signal at user $k$ is given as follows:
\begin{align}\label{received signal NOMA}
	{y_k} = \left( {{\mathbf{r}}_k^H{{\mathbf{\Theta }}_k}{\mathbf{g}}} \right)\left( {\sqrt {{p_t}} {s_t} + \sqrt {{p_r}} {s_r}} \right) + {n_k},
\end{align}
where $n_k \sim {\mathcal{C}\mathcal{N}}\left( {0,{\sigma ^2}} \right)$ denotes the additive white Gaussian noise at user $k$ with variance $\sigma^2$. Let the binary variable ${\lambda  \left( k \right) \in \left\{ {0,1} \right\}}$ denote the decoding order of user $k$, which satisfies $\lambda \left( t \right) + \lambda \left( r \right) = 1$. For instance, if the T user is the strong user with higher channel power gain (i.e., ${\left| {{h_t}} \right|^2} \ge {\left| {{h_r}} \right|^2}$), which first decodes the signal of R user before decoding its own signal, i.e., through successive interference cancellation (SIC), we have $\lambda \left(t \right)=1$ and $\lambda \left( r \right)=0$. Otherwise, $\lambda \left( t \right)=0$ and $\lambda \left( r \right)=1$\cite{mu2}. Therefore, the achievable communication rate of user $k$ for NOMA is given by
\begin{align}\label{NOMA rate}
	r_k^N = {\log _2}\left( {1 + \frac{{{\rho _0}{p_k}{\beta _k}{{\left| {{{\mathbf{q}}_k}{{\mathbf{v}}_k}} \right|}^2}}}{{\lambda \left( {\bar k} \right){\rho _0}{p_{\bar k}}{\beta _k}{{\left| {{{\mathbf{q}}_k}{{\mathbf{v}}_k}} \right|}^2} + D_k^{{\alpha _{RU}}}{\sigma ^2}}}} \right),
\end{align}
where $\bar{k}$ represents the other user.

\subsubsection{OMA} For OMA, we consider the general case where the AP transmits the signals of the two users throughout the orthogonal frequency/time resources which can be allocated adaptively\cite{OMA}. Let $\omega_k\in[0,1]$ denote the proportion of resource blocks allocated to user $k$. Then, the achievable rate of user $k$ for OMA can be expressed as
%\begin{align}\label{OMA rate1}
%r_k^O = \frac{1}{2}{\log _2}\left( {1 + \frac{{{\rho _0}{p_k}{\beta _k}{{\left| {{{\mathbf{q}}_k}{{\mathbf{v}}_k}} \right|}^2}}}{{\frac{1}{2}D_k^{{\alpha _{RU}}}{\sigma ^2}}}} \right).
%\end{align}
%If the the orthogonal frequency/time resources is adaptively allocated to users, namely OMA-II, the achievable rate of user $k$ can be expressed as
\begin{align}\label{OMA rate2}
r_k^O = {\omega _k}{\log _2}\left( {1 + \frac{{{\rho _0}{p_k}{\beta _k}{{\left| {{{\mathbf{q}}_k}{{\mathbf{v}}_k}} \right|}^2}}}{{{\omega _k}D_k^{{\alpha _{RU}}}{\sigma ^2}}}} \right).
\end{align}

\subsection{Problem Formulation}

In this article, we aim to characterize the coverage range of STAR-RIS for NOMA and OMA, subject to the predefined QoS requirements of users. %Due to the quality of service (QoS) requirements of users and the limited network resources, the STAR-RIS's coverage ranges of the transmission region and reflection region are mutually restricted. 
Let $D_0$ denote the maximum coverage range provided by STAR-RIS, which is the sum of the transmission coverage range, $D_t$, and the reflection coverage range, $D_r$. Moreover, let $\mu_k$ denote the coverage allocation factor for user $k$, where user $k$ can be served successfully within the maximum distance ${\mu_k}{D_0}$. We have $\mu_k\geq0$ and $\mu_t+\mu_r=1$. A larger value of $\mu_k$ means a higher priority of user $k$. Then, the characterization of the coverage range of the STAR-RIS for NOMA can be formulated as the following optimization problem: 

%Then, the optimization problem for characterizing coverage range for NOMA can be formulated as

%Let $\mathbf{\Upsilon}$ denotes the feasible sets of $\{p_k,\beta_k,w_k,{\mathbf{\Theta}_k}\}$ that satisfy the resource and phase shift constraint. Then, the coverage range is defined as 
%\begin{equation}\label{def1}
%	\mathcal{D}(D_k)\triangleq  \underset{\{p_k,\beta_k,w_k,{\mathbf{\Theta}_k}\}\in\mathbf{\Upsilon}}{\cup} \big\{D_k:r_k\geq\gamma_k,\forall{k}\big\}.
%\end{equation}
%
%From defination (\ref{def1}), the coverage range is the coverage distance pairs that satisfy the quality-of-service (QoS) requirement of each user. The upper-right boundary of this set characterizes the whole coverage range, at which we can only sacrifice the communication distance of one user to improve that of the other user. We further invoke a variable $D_0$ to represent the whole coverage distance and a constant $0\leq\mu\leq1$ to depict the tradeoff between the two side. Specifically, $\mu{D_0}$ is the coverage range of the transmission region while $(1-\mu)D_0$ is the coverage range of reflection region. 

\begin{subequations}\label{P1 NOMA}
	\begin{align}
		&\mathop {\max}\limits_{\left\{ {{p_k},{\beta _k},{D_k},\lambda \left( k \right),{{\mathbf{v}}_k}},{D_0} \right\}} \;\; D_0 \\
		\label{dis NOMA}{\rm{s.t.}}\;\;	&D_k\geq{u_kD_0}, \forall{k}\in \mathcal{K}, \\
		\label{far field1}&{D_k}\geq1,\forall{k}\in \mathcal{K},\\
	    %\label{QoS OMA}	&r_k^O \ge {\overline r _k},\forall k \in {\mathcal{K}},\\
	     \label{QoS NOMA}	&r_k^N \ge {\gamma_k},\forall k \in {\mathcal{K}},\\
		\label{power NOMA}&\sum_{k}p_k\leq{P_{\text{max}}}, \\
		%\label{phase shift OMA}&\left| {{{\left[ {{{\mathbf{v}}_k}} \right]}_m}} \right| = 1,\forall m \in {\mathcal{M}},k \in {\mathcal{K}},\\
		\label{phase shift NOMA}&\theta_m^k\in[0,2\pi],\forall m \in {\mathcal{M}},k \in {\mathcal{K}},\\
		\label{mode NOMA}&{\beta_r} + {\beta_t} = 1,\\
		\label{decoding order1}&\lambda \left( k \right) \in \left\{ {0,1} \right\},\lambda \left( t \right) + \lambda \left( r \right) = 1,\\
		\label{decoding order2}&
		\left\{ \begin{gathered}
			{\left| {{h_t}} \right|^2} \ge {\left| {{h_r}} \right|^2},{\rm{if}}\;\lambda \left( t \right) = 1 \hfill \\
			{\left| {{h_t}} \right|^2} \le {\left| {{h_r}} \right|^2},{\rm{otherwise}} \hfill \\
		\end{gathered}  \right.,
	\end{align}
\end{subequations}
where (\ref{far field1}) ensures that the users locate at the far-field region for STAR-RIS; (\ref{QoS NOMA}) denotes the QoS requirements of users; (\ref{power NOMA}) denotes the total power constraint; (\ref{phase shift NOMA}) is the phase shift constraint for each element of STAR-RIS; (\ref{mode NOMA}) represents the energy conservation constraint; (\ref{decoding order1}) and (\ref{decoding order2}) are the decoding order constraints for SIC. 
%Note that the different expression of $r_k^O$ in (\ref{OMA rate1}) and (\ref{OMA rate2}) represents the problem OMA-I and OMA-II.

For OMA, the optimization problem can be formulated as follows:
\begin{subequations}\label{P2 OMA}
\begin{align}
%({{\rm{P1^{OMA}}}}):
&\mathop {\max }\limits_{\left\{ {{p_k},\beta_k,{\omega_k},{D_k},{{\mathbf{v}}_k}},D_0 \right\}} \;\;{D_0} \\
{\rm{s.t.}}\;\;&
\label{QoS OMA}r_k^O \ge {\gamma_k},\forall k \in {\mathcal{K}},\\
%\label{phase shift NOMA}&\left| {{{\left[ {{{\mathbf{v}}_k}} \right]}_m}} \right| = 1,\forall m %\in {\mathcal{M}},k \in {\mathcal{K}},\\
	&\label{band OMA}\sum_{k}\omega_k\leq1,\\
%%\label{mode NOMA}&{\beta _1} + {\beta _2} = 1.
&(\rm\ref{dis NOMA}),(\rm\ref{far field1}),(\rm{\ref{power NOMA})}-(\ref{mode NOMA}).
\end{align}
\end{subequations}

The main challenges for solving the two problems are as follows: Firstly, the introduced transmission and reflection coefficients of the STAR-RIS are coupled with the variables of network resource allocation, which increases the complexity of the optimization problems. Secondly, for NOMA, the decoding order constraint for SIC has to be effectively tackled since it is non-convex.

%\begin{figure}[t!]
%  \centering
%  \includegraphics[width=5in]{eps/converge_tradeoff.eps}\\
%  \caption{Illustration of an example and the expected results.}\label{tradeoff}
%\end{figure}
\remark For the two-user multiple access scenario with STAR-RIS, ${{\mathbf{v}}_k}$ can be maximized independently by combining signals from different paths coherently. 

\rm{}It is noted that with STAR-RIS, the phase shift vectors ${\mathbf{v}}_k$ for the two sides can be optimized independently. In the two user scenario, since the achievable rate monotonically increases with $|{{\mathbf{q}}_k}{{\mathbf{v}}_k}|$, we can simply maximize them by adjusting the incident signals to have the same phases. Then, we denote $c_k=\text{max}_{\mathbf{v}_k}|{{\mathbf{q}}_k}{{\mathbf{v}}_k}|^2=\sum_{m=1}^M(|q_{m,k}|^2)$. 
\section{Proposed Solutions}
\subsection{NOMA Case}
The problem for NOMA case is complex due to the binary decoding order (\ref{decoding order1}) and non-convex  constraint (\ref{decoding order2}) for SIC. To handle these difficulties, we present the following lemma:
\begin{lemma}\label{lemma1}
	The non-convex constraint (\ref{decoding order2}) for SIC can be transformed into linear one related to reflection coefficient $\beta_k$ and rate requirement $\gamma_k$.
\end{lemma}
\newcounter{mytempeqncnt}
\setcounter{mytempeqncnt}{\value{equation}}
\begin{figure*}[!t]
	% ensure that we have normalsize text
	\normalsize
	\setcounter{equation}{8}
	% Store the current equation number.
	
	% Set the equation number to one less than the one
	% desired for the first equation here.
	% The value here will have to changed if equations
	% are added or removed prior to the place these
	% equations are referenced in the main text.
	\begin{equation}	\label{la1}
		\begin{aligned}
			\mathcal{L}_1(\mathbf{y},\boldsymbol{\mu},\lambda,\upsilon)=\upsilon(\sum_{k}p_k-P_{\text{max}})+\lambda(\sum_{k}\beta_k-1)+\mu_1\big[(2^{\gamma_t}-1)(a\frac{D_t^{\alpha_{RU}}}{\beta_t}+p_r)-{p_t}\big]  
			+\mu_2b\big[(2^{\gamma_r}-1)\frac{D_r^{\alpha_{RU}}}{\beta_r}-{p_r}\big]+E.
		\end{aligned}
	\end{equation}
	% Restore the current equation number.
	% IEEE uses as a separator
	\hrulefill
	% The spacer can be tweaked to stop underfull vboxes.
	\vspace*{4pt}
\end{figure*}
\setcounter{equation}{\value{mytempeqncnt}}

\proof We take $\lambda(r)=1$ as an example and give the Lagrangian function $\mathcal{L}_1(\mathbf{y},\boldsymbol{\mu},\lambda,\upsilon)$ which is shown in (\ref{la1}). 
%\begin{equation}
%	\begin{aligned}
%			\mathcal{L}_2(\mathbf{y},\boldsymbol{\mu},\boldsymbol{\lambda},\boldsymbol{\upsilon})=\upsilon_3(p_1+p_2-P_{\text{max}})+\upsilon_4(\beta_1+\beta_2-1)\\+\mu_1\big[(2^{\gamma_1}-1)(k_1\frac{D_1^{\alpha_{RU}}}{\beta_1}+p_2)-{p_1}\big]  
%	+\mu_2\big[(2^{\gamma_2}-1)k_2\frac{D_2^{\alpha_{RU}}}{\beta_2}-{p_2}\big]
%	\end{aligned}
%\end{equation}
where $\mathbf{y}=\{{p_k},{\beta _k},{D_k},D_0,\forall{k}\}$ denotes the variable set. The Karush-Kuhn-Tucker (KKT) conditions are listed as follows:
\setcounter{equation}{\value{equation}+1}
\begin{subequations}\label{kkt2}
	\begin{align}
		&\nabla_{\beta_k^*}\mathcal{L}_1=(2^{\gamma_k}-1)\mu_k\frac{(D_k^*)^{\alpha}}{(\beta_k^*)^2{c_k}}-\lambda=0, \\
		&\nabla_{p_t^*}\mathcal{L}_1=\upsilon-\mu_1=0,\\
		&\nabla_{p_r^*}\mathcal{L}_1=(2^{\gamma_t}-1)\mu_1-\mu_2+\upsilon=0,
	\end{align}
\end{subequations}
where $E$ denotes  the terms which are independent of $\beta_k$ and $p_k$, $\boldsymbol{\mu}=[\mu_1,\mu_2],\upsilon,\lambda$ are non-negative Lagrangian multipliers. Then, we have $(\frac{\beta_r^*}{\beta_t^*})^2=\frac{2^{\gamma_t}(2^{\gamma_r}-1)}{2^{\gamma_t}-1}(\frac{D_r}{D_t})^{\alpha}\frac{a}{b}$. Then{ $|{{h_t}|^2} \leq{|h_r|}^2 \Leftrightarrow \frac{\beta_t{a}}{D_t^{\alpha}}\geq\frac{\beta_r{b}}{D_r^{\alpha}}\Leftrightarrow{\beta_r\leq\frac{2^{\gamma_t}(2^{\gamma_r}-1)}{2^{\gamma_t}-1}}\beta_t$. Following the similar derivation, for the decoding order $\lambda(t)=1$, the non-convex constraint for SIC $|{{h_t}|^2}\geq{|h_r|}^2$  is equivalent to $\beta_t\leq\frac{2^{\gamma_r}(2^{\gamma_t}-1)}{2^{\gamma_r}-1}\beta_r$, which is a linear one. %Note that the transformed constraint is easy to meet with a specific decoding order and high rate requirements.

The relaxed problem for a given decoding order is convex, which can be proved by the following lemma:

\lemma \label{lemma2} For $x,y>0$ and $\alpha\geq2$, $f(x,y)=\frac{y^\alpha}{x}$ is a convex function with respect to $x$ and $y$.
\proof  When $x,y>0$ and $\alpha\geq2$, it is easy to get $\frac{\partial^2{f}}{\partial{x^2}}\cdot\frac{\partial^2{f}}{\partial{y^2}}-(\frac{\partial^2{f}}{\partial{x}\partial{y}})^2=(\alpha^2-2\alpha)y^{2\alpha-2}x^{-4}\geq0$ and $\frac{\partial^2{f}}{\partial{x^2}}=2y^{\alpha}x^{-3}\geq0$; then, the Hessian matrix of function $f(x,y)$ is positive semidefinite. Thus, $f(x,y)$ is convex.
%\begin{lemma}\label{lemma4}
%	For a given decoding order and neglect the constraint (\ref{decoding order2}), the relaxed problem NOMA is convex. 
%\end{lemma}

We also take the decoding order $\lambda(r)=1$ as an example and rewrite the QoS requirement constraint (\ref{QoS NOMA}). Then, the transformed problem for NOMA is given by:
{\setlength\abovedisplayskip{-10pt}
	
	\setlength\belowdisplayskip{3pt}
\begin{subequations}\label{qos noma cvx}
	\begin{align}
		&\mathop {\max}\limits_{\left\{ {{p_k},{\beta _k},{D_k}},{D_0} \right\}} \;\; D_0 \\
		&(2^{\gamma_t}-1)(a\frac{D_t^{\alpha_{RU}}}{\beta_t}+p_r)\leq{p_t}, \\
		&(2^{\gamma_r}-1)b\frac{D_r^{\alpha_{RU}}}{\beta_r}\leq{p_r},\\
		&{\beta_r\leq\frac{2^{\gamma_t}(2^{\gamma_r}-1)}{2^{\gamma_t}-1}}\beta_t,\\
		&\rm{(\ref{dis NOMA}),(\ref{far field1}),(\ref{power NOMA}),(\ref{mode NOMA}).}
	\end{align}
\end{subequations}
where $a=\frac{\sigma^2}{\rho_0c_t}$ and $b=\frac{\sigma^2}{\rho_0c_r}$ are constants with respect to environmental parameters and phase shifts. Based on \textbf{Lemma} \ref{lemma2}, it is observed that problem (11) is convex. For NOMA, the coverage characterization problem (7) can be solved by exhaustively searching over 2 decoding orders. The coverage range is then given by $D_0^*=\mathop{\arg\max}_{k\in\mathcal{K}}{(D^*_{0(\lambda(k)=1)})}$. The computational complexity of solving the NOMA problem is given by $\mathcal{O}(2N_1^{3.5})$\cite{convex}, where $N_1=7$ is the number of variables.}
\subsection{OMA Case}
%It is difficult to intuitively know whether the problem is convex for case OMA since there are multiple variables which are coupled. 
Before solving the problem, we first have the following lemma, which reveals the relation between optimal amplitude coefficient and power allocation.

\begin{lemma}\label{lemma3}
	For the case OMA, the optimal power allocation $p^{*}_k$ and amplitude coefficient $\beta^{*}_k$ satisfy $\frac{p_t^*}{\beta_t^*}=\frac{p_r^*}{\beta_r^*}=P_{\rm{max}}$.
\end{lemma}
\proof The Lagrangian function of the problem is given by
\begin{equation}
	\begin{aligned}
		\mathcal{L}_2(\mathbf{x},\boldsymbol{\mu},\lambda,\upsilon)=\mu_k[{\omega _k}{\log _2}\left( {1 + \frac{{{\rho _0}{p_k}{\beta _k}c_k}}{{{\omega _k}D_k^{{\alpha _{RU}}}{\sigma ^2}}}} \right)-\gamma_k]\\+\upsilon(p_t+p_r-P_{\text{max}})+\lambda(\beta_t+\beta_r-1)+E,
	\end{aligned}
\end{equation}
where $\mathbf{x}=\{{p_k},{\beta _k},{D_k},{\omega_k},D_0,\forall{k}\}$ is the variable set, $E$ denotes the terms which are independent of $\beta_k$ and $p_k$, $\boldsymbol{\mu}=[\mu_1,\mu_2]$, $\upsilon$ are non-negative Lagrangian multipliers, and $\lambda$ is multiplier regarding the equality constraint. Since the KKT conditions are necessary for all local optimal points, we have $\nabla_{p_k^*}\mathcal{L}_2=0$ and $\nabla_{\beta_k^*}\mathcal{L}_2=0$ which are expressed as:
\begin{subequations} \label{proof2}
	\begin{align}
		&\nabla_{\beta_k^*}\mathcal{L}_2=-\log_2{(e)}\mu_k{\omega_k^*}\frac{ap_k^*}{1+ap_k^*{\beta_k^*}}+\upsilon=0 ,\\
		&\nabla_{p_k^*}\mathcal{L}_2=-\log_2{(e)}\mu_k{\omega_k^*}\frac{a\beta_k^*}{1+ap_k^*{\beta_k^*}}+\lambda=0,
	\end{align}
\end{subequations}
where $a=\frac{{{\rho_0}{c_t}}}{{{\omega_t^*}(D_t^*)^\alpha{\sigma ^2}}}$, $b=\frac{{{\rho_0}{c_r}}}{{{\omega_r^*}(D_r^*)^\alpha{\sigma ^2}}}$, and $\alpha=\alpha_{RU}$. With (\ref{proof2}), we get $\frac{p_t^*}{\beta_t^*}=\frac{p_r^*}{\beta_r^*}=\frac{\upsilon}{\lambda}$. It is easy to deduce that $\upsilon>0$ for $\gamma_k>0$. With the complementary slackness conditions $\upsilon(p_t^*+p_r^*-{P_{\text{max}}})=0$, constraint (\ref{power NOMA}) is active and can be met with equality. The linear relation of $p_k^*$ and $\beta_k^*$ is proved and the ratio between them is $P_{\text{max}}$. \textbf{Lemma} \ref{lemma3} gives us an interesting insight to reduce the number of optimization variables. Furthermore, based on \textbf{Lemma} \ref{lemma2} and \textbf{Lemma} \ref{lemma3}, we find that with fixed time/bandwidth allocation $\omega_k=\omega_0$, the 
subproblem for OMA is convex, which is expressed as:
%rate requirement constraints (\ref{QoS OMA}) is convex which is equivalent to:

%Then, we can conclude that the optimal power allocation and amplitude coefficient have the same tendency. We can use methods like successive convex approximation to find the local optimal solution by treating $\beta_k{p_k}$ as one variable. However, we further find that with fixed time/bandwidth allocation $w_k=w_0$, the rate requirement constraints (\ref{QoS OMA}) is convex which is equivalent to:
\begin{subequations}
	\begin{align}
	%({{\rm{P2^{OMA}}}}):
	&\mathop {\max }\limits_{\left\{ {\beta_k,{D_k}},D_0 \right\}} \;\;{D_0} \\
	&\frac{\omega_0(2^{{\frac{\gamma_k}{\omega_0}}}-1)\sigma^2}{\rho_0{c_k}}\frac{D_k^{\alpha_{RU}}}{\beta_k}-P_{\rm{max}}\beta_k\leq0,\\
	&\rm{(\ref{dis NOMA}),(\ref{far field1}),(\ref{mode NOMA}).}
	\end{align}
\end{subequations}

Problem (14) is a convex optimization problem, which can be solved efficiently by solvers like CVX\cite{cvx}. Thus, the optimal solution for problem (8) can be obtained using the one-dimensional search over $0\leq{\omega_k}\leq1$. The computational complexity of solving the OMA problem is given by $\mathcal{O}(N_2^{3.5}\log_2(1/\epsilon))$\cite{convex}, where $N_2=5$ is the number of variables and $\epsilon$ is the accuracy for one-dimensional search.

\section{Simulation Results}
\begin{figure}[t]
	\centering
	\includegraphics[width=0.45\textwidth]{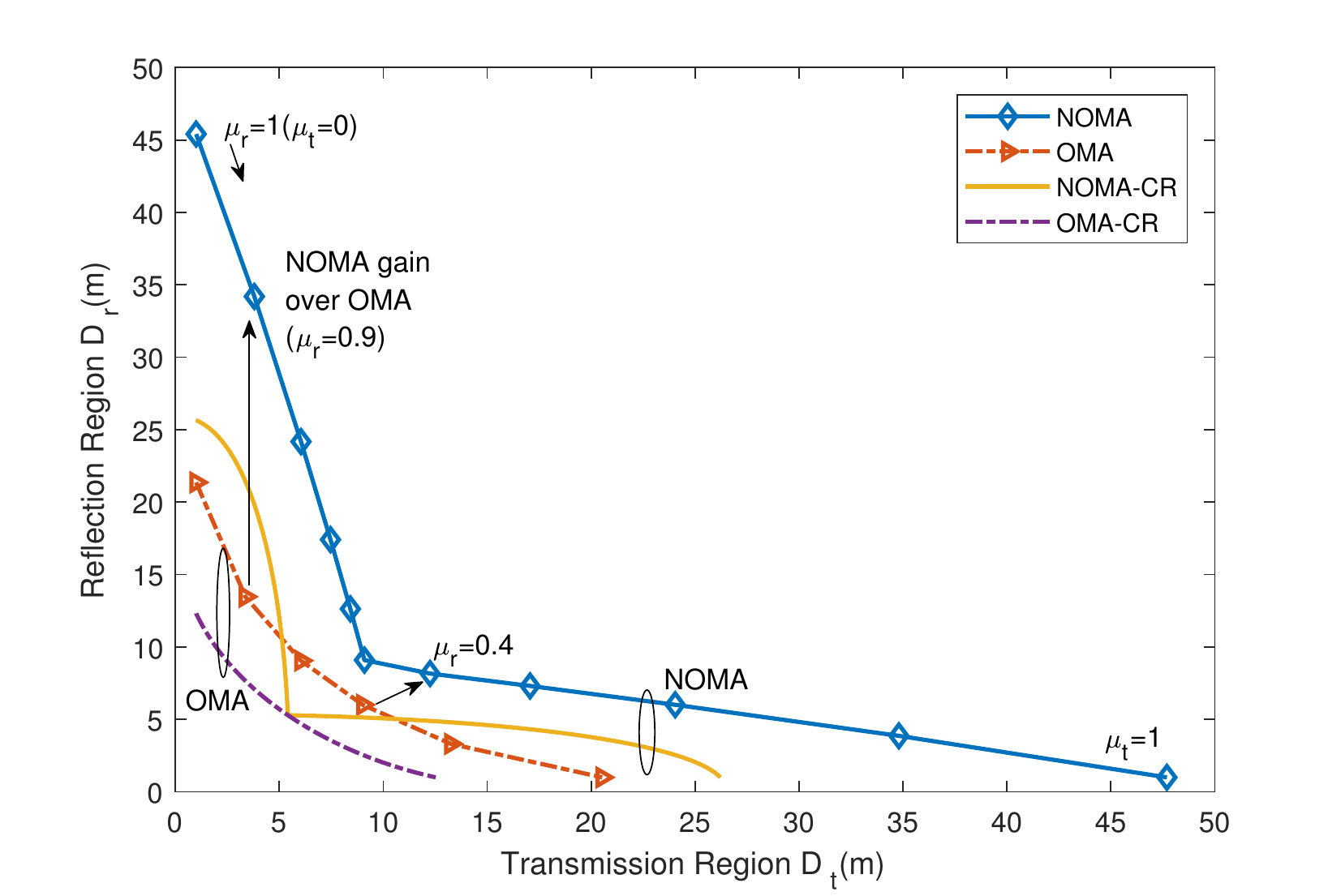}\\
	\caption{The coverage range $D_t$ and $D_r$ with QoS requirements $\gamma_k=$5 bps/Hz, where "CR" refers to conventional RISs.}
\end{figure}

%\begin{figure}[t]
%	\centering
%	\subfloat[The coverage distance $D_1$ and $D_2$ with common rate requirements 5 bps/Hz]	{
%		\includegraphics[height=3.3cm,width=4.4cm]{eps/pareto.eps}	
%	}
%	\subfloat[The coverage distance $D_1$ versus $\gamma_1$, with fixed user requirements $\gamma_1+\gamma_2=10$ bps/Hz.]{	
%		\includegraphics[height=3.3cm,width=4.4cm]{eps/sum_rate.eps}
%	}	
%	\caption{Performance comparison with fixed coefficient.}
%\end{figure}
In this section, we provide numerical results to validate the effectiveness of our proposed designs. We consider a two-dimensional coordinate system, where the STAR-RIS is equipped with $M=100$ elements. The AP is located at the origin, while the STAR-RIS is located at (50,0) meters. We set $\rho_0=-30$ dB, $\sigma^2=-80$ dBm, ${{\alpha_{RU}}}=\alpha_{AR}=2.2$, $K_{RU}=K_{AR}=10$, $P_{\text{max}}=30$ dBm. 

For performance comparison, we consider conventional RISs (CR) as a benchmark scheme. In particular, we employ one reflecting-only RIS and one transmitting-only RIS, each consisting of $M$/2 elements to achieve full-space coverage for a fair comparison with the STAR-RIS.

%\item a) OMA-TYPE-II\cite{OMA}: It represents OMA systems with optimum power allocation and fixed time/frequency allocation ($w_k=0.5$), which is a special case of OMA.
%a) Fixed transmission and reflection coefficient ($\beta=0.5$): Here we set as a constant to represent a special case, where the incident signals are always equally divided. 
%\item b) Conventional RIS: We employ one reflecting-only RIS and one transmitting-only RIS, each of which has $M$/2 elements to achieve full-space coverage. For a fair comparison, the total number of elements is consistent with the STAR-RIS.

%\begin{figure}[t]
%	\centering
%	\includegraphics[width=0.4\textwidth]{eps/system_setup.eps}\\
%	\caption{Simulation setup}\label{setup}
%\end{figure}

In Fig. 2, we characterize the coverage range of the STAR-RIS with different coverage range allocation factors ${\mu_k}$. We set the QoS requirements as $\gamma_k=5$ bps/Hz and plot the coverage range pairs $(D_t,D_r)$. By varying $\mu_k$, we get different tuples of the optimal coverage range for the transmission region and reflection region. %These boundaries characterize the whole coverage range, at which we can only sacrifice the communication distance of one user to improve that of the other user. 
It is observed that for NOMA, the performance gain in terms of coverage range is not prominent for homogenous priority, i.e., $\mu_t$ and $\mu_r$ are close. The performance gain of NOMA over OMA is more pronounced as $\mu_k$ approaches 0 or 1, since the multiplexing gain of the time/frequency resource is more remarkable. It is also observed that by deploying the STAR-RIS, the coverage range is nearly doubled for both NOMA and OMA compared with that of conventional RISs, which verifies the effectiveness of the proposed STAR-RIS.
%since STAR-RIS elements work at both side and the channel gain from STAR-RIS to user is larger.
%It is also worth mentioning that, for most of time, OMA outperforms OMA-TYPE-II since it allows more flexible adjustment of spectrum or time resources according to users' requirements. Moreover, the coverage range of STAR-RIS completely covers that of conventional RIS.

In Fig. 3, we depict the total coverage range, $D_0$, versus different T user's QoS requirement, $\gamma_t$. We fix the R user's QoS requirements as $\gamma_r=5$ bps/Hz and the coverage allocation factor $\mu_t=0.6$. %since the channel condition of the reflection region approximates that of the transmission region, the performance gain of NOMA over OMA for conventional RIS is negligible. 
As seen from Fig. 3, for conventional RISs, the performance gap between NOMA and OMA is negligible.
However, for STAR-RIS, NOMA achieves a significant performance gain in terms of coverage range compared with OMA, i.e., \emph{NOMA gain}. This is because the simultaneously transmitting and reflecting scheme can enlarge the channel disparity between the two users, where NOMA yields higher performance gain than OMA. Furthermore, the STAR-RIS provides a significant performance gain for both NOMA and OMA compared with conventional RIS, i.e., \emph{STAR gain}. The results also confirm the superiority of the proposed STAR-RIS.
%\begin{figure}[H]
%	\centering
%	\includegraphics[width=0.4\textwidth]{eps/fixbeta.png}\\
%	\caption{Coverage distance versus $\mu$ with fixed $\beta$ }\label{3}
%\end{figure}

In Fig. 4, we plot the coverage range of the T user, $D_t$, versus the number of elements, $M$. The QoS requirements are set to $\gamma_k=3$ bps/Hz and the coverage allocation factor is $\mu_t=0.6$. It can be observed that the coverage range increases linearly as there are larger total number of elements for all cases. The STAR gain over conventional RISs is more pronounced as $M$ increases.
%Despite that the incident signals are divided by the STAR-RIS, the channel gain of STAR-RIS-user is much larger than that of RIS-user. Therefore, the total coverage range is apparently extended by the STAR-RIS compared with two independently working RISs. 
Meanwhile, NOMA can also enhance the network performance of both STAR-RISs and conventional RISs for asymmetric channels, i.e., $\mu_t$ is not close to $\mu_r$. It can be seen from Fig. 4 that the NOMA gain for STAR-RISs is larger than that for conventional RISs. The results indicate that the combination of NOMA and STAR-RIS is a win-win strategy.

\begin{figure}[t]
	\centering
	\includegraphics[width=0.45\textwidth]{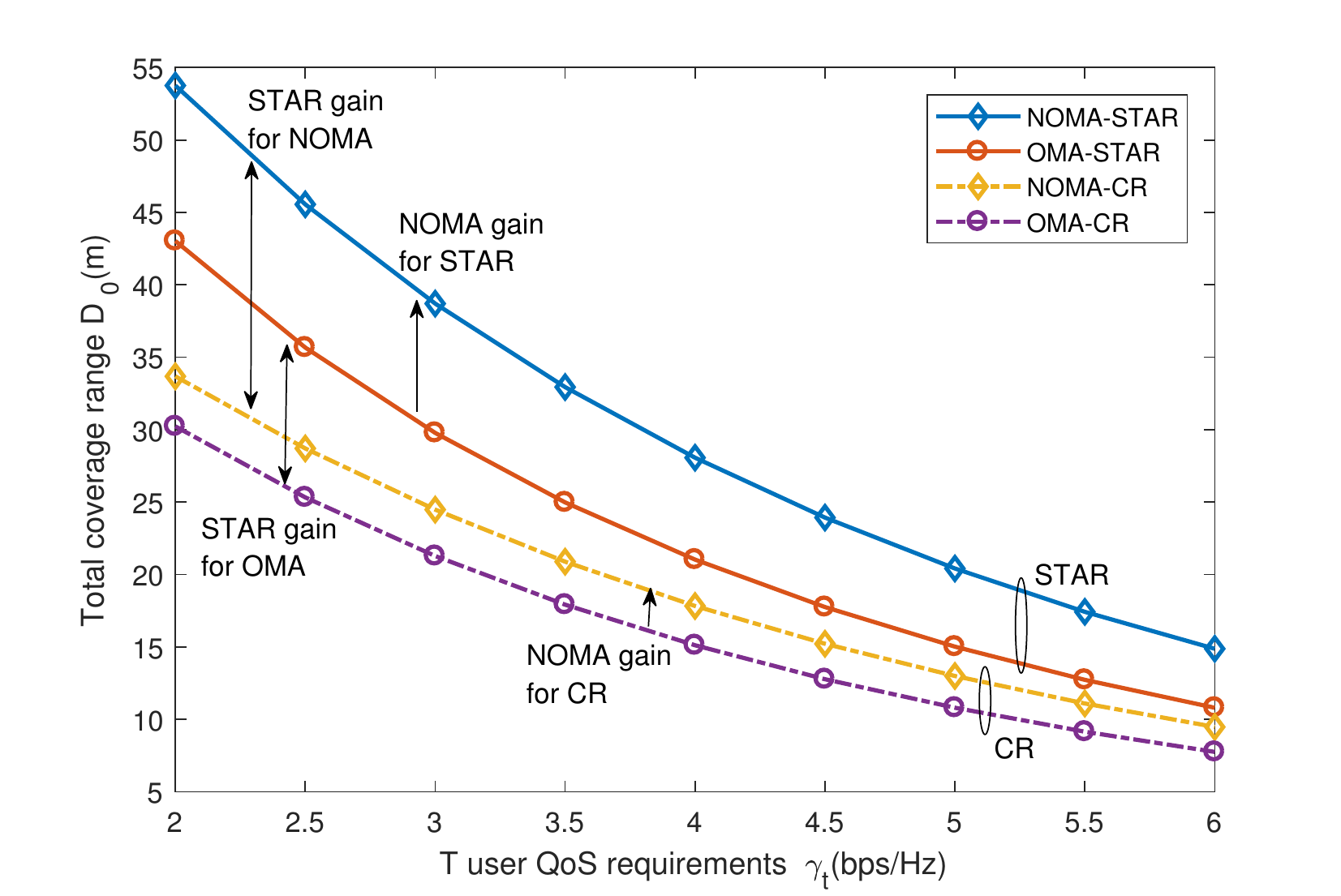}\\
	\caption{The total coverage range $D_0$ versus different $\gamma_t$, with fixed R user QoS requirements $\gamma_r=5$ bps/Hz.}
\end{figure}

\begin{figure}[t]
	\centering
	\includegraphics[width=0.45\textwidth]{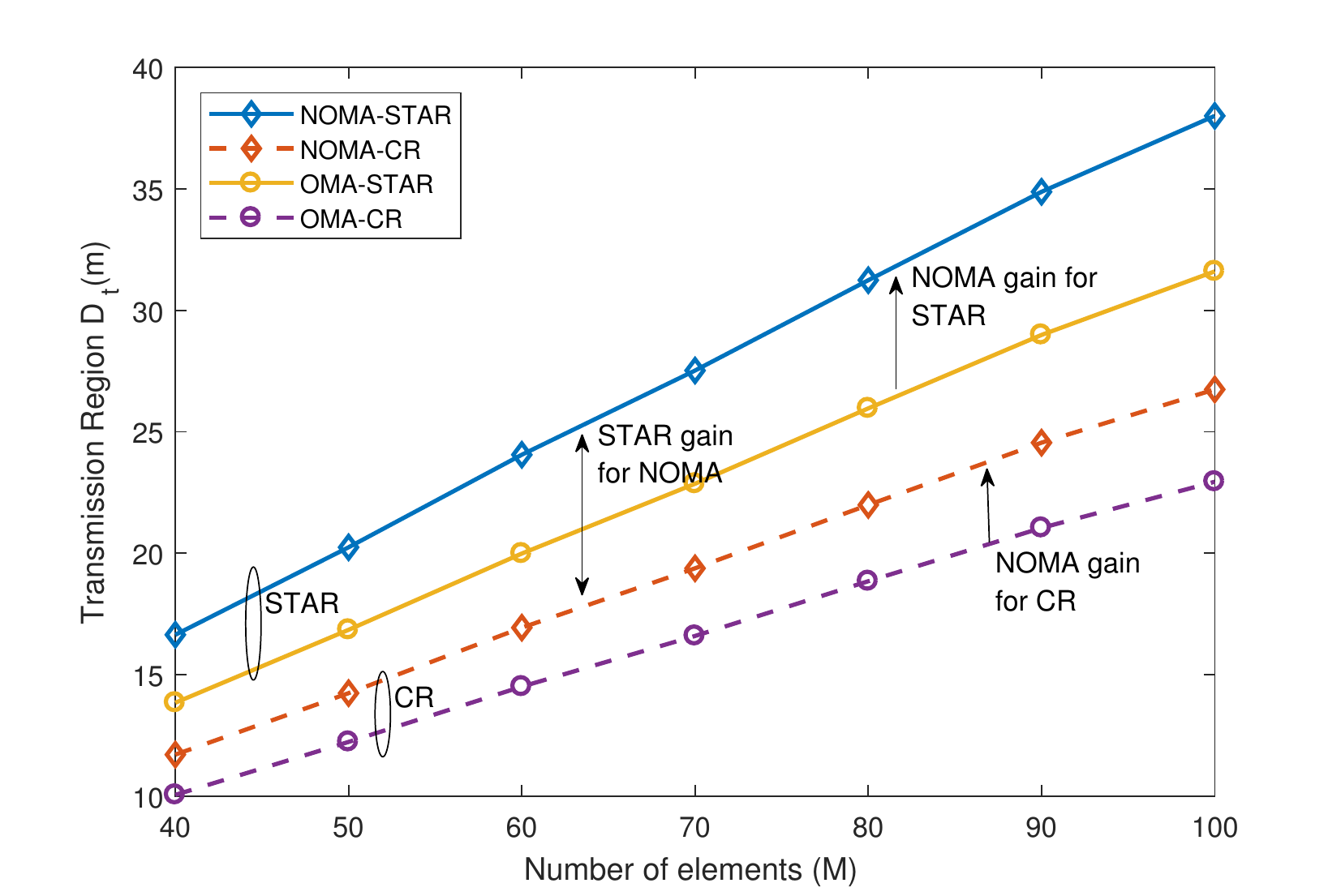}\\
	\caption{The coverage range $D_t$ versus different number of total elements, with QoS requirements $\gamma_k=3$ bps/Hz.}
\end{figure}

\section{Conclusions}
In this article, we studied the fundamental coverage characterization of STAR-RIS assisted two-user communication networks. A sum coverage range maximization problem for NOMA and OMA was formulated, which involved a joint optimization of the transmission and reflection coefficients at the STAR-RIS and resource allocation at the AP. For NOMA, we transformed the non-convex constraints to make the problem convex. For OMA, we used the one-dimensional search based algorithm to find the optimal solution. Simulation results showed that the STAR-RIS provided wider coverage range compared with conventional reflecting and transmitting-only RIS. Furthermore, the interplay between STAR-RIS and NOMA increased the flexibility for network design.
%The rate boundary with fixed user location is characterized in Fig. \ref{ratebound}.
%\begin{figure}[htbp]
%	\centering
%	\includegraphics[width=0.4\textwidth]{eps/6.eps}\\
%	\caption{The Pareto boundary of $r1$, $r2$ with fixed user location}\label{ratebound}
%\end{figure}

\bibliographystyle{ieeetr}
\bibliography{reference2.bib}

 \end{document}